\RequirePackage{ifpdf}
\ifpdf 
\documentclass[pdftex]{sigma}
\else
\documentclass{sigma}
\fi

\begin{document}
\allowdisplaybreaks

\renewcommand{\PaperNumber}{059}

\FirstPageHeading

\ShortArticleName{Finding Liouvillian First Integrals of Rational
ODEs of Any Order in Finite Terms}

\ArticleName{Finding Liouvillian First Integrals\\ of Rational
ODEs of Any Order in Finite Terms}

\Author{Yuri N. KOSOVTSOV} 
\AuthorNameForHeading{Yu.N. Kosovtsov}

\Address{Lviv Radio Engineering Research Institute, 7 Naukova Str., 
Lviv, 79060 Ukraine}
\Email{\href{mailto:kosovtsov@escort.lviv.net}{kosovtsov@escort.lviv.net}}

\ArticleDates{Received August 31, 2005, in f\/inal form May 12,
2006; Published online June 08, 2006}

\Abstract{It is known, due to Mordukhai-Boltovski, Ritt, Prelle,
Singer, Christopher and others, that if a given rational ODE has a
Liouvillian f\/irst integral then the corresponding integrating
factor of the ODE must be of a very special form of a product of
powers and exponents of irreducible polynomials. These results
lead to a partial algorithm for f\/inding Liouvillian f\/irst
integrals. However, there are two main complications on the way to
obtaining polynomials in the integrating factor form. First of
all, one has to f\/ind an upper bound for the degrees of the
polynomials in the product above, an unsolved problem, and then
the set of coef\/f\/icients for each of the polynomials by the
computationally-intensive method of undetermined parameters. As a
result, this approach was implemented in CAS only for f\/irst and
relatively simple second order ODEs. We propose an algebraic
method for f\/inding polynomials of the integrating factors for
rational ODEs of any order, based on examination of the resultants
of the polynomials in the numerator and the denominator of the
right-hand side of such equation. If both the numerator and the
denominator of the right-hand side of such ODE are not constants,
the method can determine in f\/inite terms an explicit expression
of an integrating factor if the ODE permits integrating factors of
the above mentioned form and then the Liouvillian f\/irst
integral. The tests of this procedure based on the proposed
method, implemented in Maple in the case of rational integrating
factors, conf\/irm the consistence and ef\/f\/iciency of the
method.}


\Classification{34A05; 34A34; 34A35}

\section[The Prelle-Singer method]{The Prelle--Singer method}

$\mu$  is an integrating factor for an $n$th order ODE in solved
form
\begin{gather}
\frac{d^ny_0}{dx^n}-f(x,y_0,y_1,\dots,y_{n-1})=0, \label{ode}
\end{gather}
where  $y_0 = y(x)$,    $y_j = \frac{d^jy(x)}{dx^j}$, by standard
def\/inition~\cite {Olver}, if $\mu(y_n-f)$  is a total derivative
of some function $\zeta(x,y_0,y_1, \dots , y_{n-1})$, that is,
\begin{gather}
\mu\left(\frac{d^ny_0}{dx^n}-f\right) = \frac{\partial
\zeta}{\partial x}+\sum_{j=0}^{n-1}y_{j+1}\frac{\partial
\zeta}{\partial y_j}  . \label{2}
\end{gather}
Let
\begin{gather}
\mu = \frac{\partial \zeta}{\partial y_{n-1}}. \label{mu}
\end{gather}
Then (\ref{2}) becomes
\begin{gather}
 D(\zeta)+f\frac{\partial \zeta}{\partial y_{n-1}} = 0,
\label{pde}
\end{gather}
where
\begin{gather*}
D = \frac{\partial } {\partial x}+\sum_{j=0}^{n-2}
y_{j+1}\frac{\partial }{\partial y_j}. 
\end{gather*}
We conclude that the integrating factor of ODE (\ref{ode}) is $\mu
= \frac{\partial \zeta}{\partial y_{n-1}}$, where the function
$\zeta$  is \emph{a first integral}, a solution of the linear
f\/irst-order PDE (\ref{pde}).

If we eliminate the function $\zeta$ from the system (\ref{mu}),
(\ref{pde}), we obtain a PDE system for the integrating factor.
\emph{One} of the equations of the PDE system obtained by double
dif\/ferentiation of (\ref{pde}) with respect to $y_{n-1}$ is as
follows ($n\geq2$):
\begin{gather}
D\left(\frac{\partial \mu}{ \partial
y_{n-1}}\right)+\frac{\partial^2 (f\mu)}{
\partial y_{n-1}^2}+2\frac{\partial \mu}{ \partial y_{n-2}}=0.
\label{must}
\end{gather}

Assuming that an integrating factor is known, it is well-known
that we are able to obtain corresponding f\/irst integral by
quadratures. If we know $n$ independent f\/irst integrals for
ODE~(\ref{ode}), we can then f\/ind its general solution, at least
in implicit form. As it is seen from (\ref{mu}), the algebraic
structure of the integrating factor is simpler than the structure
of the f\/irst integral, so in some cases f\/inding the
integrating factors is an easier problem.

The most beautiful way of f\/inding integrating factors is the
Darboux method and its ref\/i\-ne\-ments. If $f$ in (\ref{ode}) is a
rational function with respect to all variables
$x,y_0,y_1,\dots,y_{n-1}$, we will call such an ODE a rational
one. In 1878, Darboux \cite{Darboux} proved that if a rational ODE
has at least $m(m+1)/2$, where $m=\max({\rm order}({\rm
numerator}(f),{\rm order}({\rm denominator}(f))$ invariant
algeb\-raic curves $P_i$, now known as Darboux polynomials, then
it has a f\/irst integral or an integrating factor of the form
$\prod_i P_i^{a_i}$ for suitable complex constants $a_i$. So, in
principle, the problem of f\/inding an integrating factor is
reduced here to the problem of f\/inding Darboux polynomials.

In 1983, Prelle and Singer \cite {Singer1} proved the essential
theorem that \emph{all first order} ODEs, which possess
\emph{elementary} f\/irst integrals, have an integrating factor of
the above mentioned form $\prod_i P_i^{a_i}$. Their result
unif\/ies and generalizes a number of results originally due to
Mordukhai-Boltovski~\cite{MB}, Ritt \cite{Ritt} and others. Singer
in \cite{Singer2} ref\/ines this result for ODEs, which possess
\emph{Liouvillian} f\/irst integral.

The following theorem is a corollary of the Proposition 2.2
\cite{Singer2} and of the def\/inition of the integrating factor
(\ref{mu}) (see also \cite{Christopher}):

\begin{theorem}\label{Prelle--Singer}
If the $n$-th order ODE
\begin{gather}
\frac{d^ny_0}{dx^n}=\frac{A(x,y_0,y_1,\dots,y_{n-1})}{B(x,y_0,y_1,\dots,y_{n-1})},
\label{ode2}
\end{gather}
where $A, B \in K[x,y_0,y_1,\dots,y_{n-1}]$ are polynomials, has a
local Liouvillian first integral, then there exists an integrating
factor of ODE \eqref{ode2} of the form
\begin{gather}
\mu=\prod_i P_i^{a_i} \exp\left(b_0\prod_j Q_j^{b_j}\right),
\label{mu1}
\end{gather}
where $P_i$'s, $Q_j$'s ${} \in K[x,y_0,y_1,\dots,y_{n-1}]$ are
irreducible polynomials, $a_i$'s, $b_0$ are constants and $b_j$'s
are integers.
\end{theorem}

So, to f\/ind an integrating factor to ODE (\ref{ode2}) in the
conditions of Theorem \ref{Prelle--Singer}, we f\/irst should
obtain the sets of polynomials $P_i$ and $Q_j$,  and if we succeed
in this task then we can obtain the constants $a_i$'s and $b_j$'s
from the full PDE system for the integrating factor.

To consider the ways to f\/ind the sets of polynomials $P_i$ and
$Q_j$, let us substitute the integrating factor in the form
(\ref{mu1}) into (\ref{must}). After some rearrangement, we arrive
at
\begin{gather*}
L_1B=2A\left(\frac{\partial B}{\partial y_{n-1}}\right)^2\prod_i
P_i^2\prod_j Q_j^2,
\end{gather*}
where $L_1$ is a polynomial. As $A$ and $B$ are relatively prime
polynomials, we conclude that $B$~divides one of the polynomials
$P_i$ or $Q_j$. Suppose that
\begin{gather*}
A=A_1(y_{n-1})A_2(x,y_0,y_1,\dots,y_{n-1}),\\
B=B_1(y_{n-1})B_2(x,y_0,y_1,\dots,y_{n-2})B_3(x,y_0,y_1,\dots,y_{n-1})
\end{gather*}
and
\[
\mu=A_1^{\varepsilon_0}B_1^{\varepsilon_1}B_2^{\varepsilon_2}B_3^{\varepsilon_3}\prod_i
P_i^{a_i}
\exp\left(b_0A_1^{\eta_0}B_1^{\eta_1}B_2^{\eta_2}B_3^{\eta_3}\prod_j
Q_j^{b_j}\right).
\]
Substituting these expressions for $A$, $B$ and $\mu$ into
(\ref{must}), we can conclude, by considering divisibility of
polynomials, that it is necessary that $\varepsilon_3=1$ and
$\eta_3=0$. That is, if $\frac{\partial P_i}{\partial
y_{n-1}}\equiv0$ or $D(P_i)\equiv0$ ($\frac{\partial Q_j}{\partial
y_{n-1}}\equiv0$ or $D(Q_j)\equiv0$) then such polynomials are
factors of $A$ or $B$, and they can be easily \emph{selected} and
\emph{included} into the set of candidates for the integrating
factor structure.

Without loss of generality, we will consider an integrating factor
of the form
\begin{gather}
\mu=B\prod_i P_i^{a_i} \exp\left(b_0\prod_j Q_j^{b_j}\right).
\label{mu2}
\end{gather}
and will focus our attention on the cases when $\frac{\partial
P_i}{\partial y_{n-1}}\neq0$ and $D(P_i)\neq0$ ($\frac{\partial
Q_j}{\partial y_{n-1}}\neq0$ and $D(Q_j)\neq0$).

For the case of $a_k\neq1$, substitution of (\ref{mu2}) into
(\ref{must}) leads to
\begin{gather}
L_2P_k =a_k(a_k-1) \frac{\partial P_k}{\partial
y_{n-1}}\prod_{\substack{i\\i\neq k}} P_i^2\prod_j
Q_j^2\left[BD(P_k)+A\frac{\partial P_k}{\partial y_{n-1}}\right]
\label{C1}
\end{gather}
and similarly for the case of $b_k<0$
\begin{gather}
L_3 Q_k = b_k^2\frac{\partial Q_k}{\partial y_{n-1}}\prod_i
P_i^2\prod_{\substack{j\\j\neq k}}
Q_j^{2(b_j+1)}\left[BD(Q_k)+A\frac{\partial Q_k}{\partial
y_{n-1}}\right],
 \label{C2}
\end{gather}
where $L_2$ and $L_3$ are polynomials.

All components involved in (\ref{C1}) and (\ref{C2}) are
polynomials. We assume that $P_i$ and $Q_j$ are irreducible,
relatively prime polynomials, so ${\rm gcd}\big(P_i,\frac{\partial
P_i}{\partial y_{n-1}}\big)$ and ${\rm gcd}\big(Q_i,\frac{\partial
Q_i}{\partial y_{n-1}}\big)$ are constants, where greatest common divisor is denoted by gcd. We can conclude that
($f\mid g$ means that $f$ divides $g$)
\begin{gather}
P_i \mid BD(P_i)+A\frac{\partial P_i}{\partial y_{n-1}}
\label{C12}
\end{gather}
and
\begin{gather}
Q_j \mid BD(Q_j)+A\frac{\partial Q_j}{\partial y_{n-1}}.
 \label{C22}
\end{gather}
The basic signif\/icance of relations (\ref{C12}) and (\ref{C22})
is that we can investigate each of the polynomials~$P_i$ or~$Q_j$
\emph{independently}. Properties (\ref{C12}) and (\ref{C22})
constitute the basis of the so-called Prelle--Singer method, which
originates from Darboux \cite {Darboux}, to obtain the integrating
factors and f\/irst integrals of ODEs of type (\ref{ode2}).
Supposing that
\begin{gather}
P_k = \sum_{i=0}^N \sum_{j_0=0}^N \cdots \sum_{j_{n-1}=0}^N
c_{kij_0 \cdots j_{n-1}} x^i y_0^{j_0} \cdots y_{n-1}^{j_{n-1}}
\label{Hmu}
\end{gather}
and substituting $P_k$ into (\ref{C12}), we can in principle
obtain $c_{kij_0 \cdots j_{n-1}}$'s and as a result $P_k$ (or
similarly $Q_k$) in explicit form.

To do so, we f\/irst have to establish an upper bound $N$ for the
degrees of the polynomial $P_i$ (or $Q_k$). It is known that $N$
exists, but there is no constructive method to evaluate it so far.
It is the principal weakness of the Prelle--Singer method. In the
existing implementations of the Prelle--Singer method, an upper
bound $N$ is predef\/ined by a user of the method.

With an increase of equation order the number of undetermined
variables $c_{kij_0 \cdots j_{n-1}}$ and, correspondingly, the
dimension of the system of algebraic equations for these constants
increases. As a result, solving of this system becomes
problematical. This approach was implemented in CAS only for
f\/irst and relatively \emph{simple second} order ODEs
\cite{MacCallum,Man1,Duarte1,Duarte2,
Shtokhamer,Man2,Duarte3,Duarte4}, and as a rule the method can not
be used for $N>4$. In contrast, the modif\/ied method proposed in
the next section does not suf\/fer from these restrictions.

We  must also mention that a rational ODE does not always have a
local Liouvillian f\/irst integral. While generalizations of the
Darboux integrability theory allow in some (``integrable") cases
to f\/ind non-Liouvillian f\/irst integrals (see,
e.g.~\cite{Kosovtsov02/1,Kosovtsov02/2,Gine1,Gine2,Gine3}),  we
consider only Liouvillian f\/irst integral theory.

\section{The method of resultants}

Let us investigate closely the relation (\ref{C12}) or
equivalently (\ref{C22}). Let $R_z(f,g)$ be the resultant of
polynomials $f$ and $g$ with respect to the indeterminate $z$. In
the sequel, we will consider only the case when both $A$ and $B$
are not constants and there are such $z\in
(x,y_0,y_1,\dots,y_{n-1})$ that $R_z(A,B)$ is not a constant.

The following Theorem \ref{ResAB} and Corollary are our main
results:
\begin{theorem} \label{ResAB}
If a polynomial $P_i$ satisfies the condition \eqref{C12} then for
any indeterminate $z\in (x,y_0,y_1,\dots,y_{n-1})$ the resultant
$R_z(P_i,B)$ must divide the resultant $R_z(A,B)$.
\end{theorem}

\begin{proof} We may write $P_i\,L= BD(P_i)+A\frac{\partial
P_i}{\partial y_{n-1}}$ for some polynomial $L$. Taking resultants
of both sides of this equation with $B$ and with respect to an
indeterminate $z\in (x,y_0,y_1,\dots,y_{n-1})$, we get the
following polynomial equation
\begin{gather*}
R_z(P_i,B)R_z(L,B)=R_z(A,B)R_z\left(\frac{\partial P_i}{\partial
y_{n-1}},B\right). 
\end{gather*}
As $P_i$ and $\frac{\partial P_i}{\partial y_{n-1}}$ have no
common roots, $R_z(P_i,B)$ does not divide $R_z\big(\frac{\partial
P_i}{\partial y_{n-1}},B\big)$. Therefore,
\begin{gather}
R_z(P_i,B) \mid R_z(A,B)  \label{R1}
\end{gather}
and it is obvious that we can choose $R_z(P_i,B)$ as a divisor of
$R_z(A,B)$ up to a constant fac\-tor.
\end{proof}

How can we recover $P_i$ if we know the resultant $R_z(P_i,B)$?
First of all, we note that the case when $A/B$ depends on only one
indeterminate is trivial. As we have mentioned above, if~$P_i$
depends only on one indeterminate, then such polynomials are
factors of $A$ or $B$ and we have already included them into the
set of candidates. So, we assume that $P_i$ depends at least on
two indeterminates and there exists such $z$ that $R_z(P_i,B)$ is
a \emph{polynomial}. We do not insist that $R_z(P_i,B)$ is
irreducible.

It is known from elementary facts about resultants that there
exist such polynomials $\alpha_i, \beta_i \in
K[x,y_0,y_1,\dots,y_{n-1}]$ that $\beta_i\,P_i = \alpha_i\,B+
R_z(P_i,B)$.

As a corollary of Theorem \ref{ResAB}, we have the following
result. Assuming that $P_i$'s (and $Q_j$'s) are irreducible
polynomials we conclude that

\begin{corollary*}
If polynomials $P_i$ (and respectively $Q_j$) satisfy \eqref{R1}
then the following holds
\begin{gather}
P_{i\,{\rm hyp}} = \alpha_i\,B+ c_i\, [\mbox{\rm one of
nonconstant factors of}\,\, R_z(A,B)], \label{PAB}
\end{gather}
where  $P_{i\,{\rm hyp}}=\beta_i\,P_i$, $\beta_i$ is some
polynomial, $\alpha_i\in (-1,0,1)$ and $c_i$ are constants.
\end{corollary*}

\begin{proof} If $\alpha_i\neq 0$ identically, then for any $z\in
(x,y_0,y_1,\dots,y_{n-1})$ from
\begin{gather*}
R_z(\beta_i P_i-\alpha_iB,P_i)=\gamma
R_z(\alpha_i,P_i)R_z(B,P_i)\\
\qquad = c_iR_z([\mbox{\rm one of nonconstant factors of}\,\,
R_z(A,B)],B)=R_z(B,P_i),
\end{gather*}
where $\gamma \in (-1,1)$, we conclude that
$R_z(\alpha_i,P_i)=1/\gamma$. So as $P_i$ depends at least on two
indeterminates then $\alpha_i$ must be a constant and $\alpha_i\in
(-1,0,1)$.

Now we are able to modify the Prelle--Singer method by replacing
the hypothesis (\ref{Hmu}) with~(\ref{PAB}) and f\/inding
constants $\alpha_i$ and $c_i$ from the following equation
\[
R_z\left(BD(P_{i\,{\rm hyp}})+A\frac{\partial P_{i\,{\rm
hyp}}}{\partial y_{n-1}},P_{i\,{\rm hyp}}\right)=0.
\] Then $P_i$ is a
irreducible factor of $P_{i\,{\rm hyp}}$ which satisf\/ies
(\ref{C12}).

Thus, we can use algebraic, not dif\/ferential, relations for
obtaining polynomials as in the original Darboux (and
Prelle--Singer) approach. First of all, we \emph{bypass} the
principal weakness of the Prelle--Singer method as we do not need
to establish an upper bound $N$ for the degrees of polynomial
$P_i$ here. Second, we have the advantage that f\/inding only two
undetermined constants $\alpha_i$ and $c_i$ in the $P_{i\,{\rm
hyp}}$ structure is much easier than f\/inding the set of
$c_{kij_0 \cdots j_{n-1}}$'s.
\end{proof}

\begin{example*} Let us consider a relatively simple second order
ODE
\[
y''=\frac{y'^3+y'(x-2)-y}{y'^2+(2y'-1)(x+y)-x}.
\]
For this ODE, $R_x(A,B)=-(2y_1^2+y_1-2)(y_1^2-2y_1-y_0)$ and we
can form the following hypothesis $P_{\rm
hyp}=\alpha(y_1^2+(2y_1-1)(x+y_0)-x)+c(y_1^2-2y_1-y_0)$.
Calculation of the following resultant leads to $R_{y_1}(BD(P_{\rm
hyp})+A\frac{\partial P_{\rm hyp}}{\partial y_1},P_{\rm
hyp})=\alpha^2(\alpha-c)L=0$, where $L$ is some polynomial. So,
there are two cases, when $\alpha=0$ then $P_{1\,{\rm
hyp}}=c(y_1^2-2y_1-y_0)$, and when $\alpha =c$ then $P_{2\,{\rm
hyp}}=2c(y_1-1)(y_1+y_0+x)$.

By  substitution of the irreducible factors of $P_{1\,{\rm hyp}}$
and $P_{2\,{\rm hyp}}$ to (\ref{C12}), we select the following
candidates for the integrating factor structure:
$P_1=(y_1^2-2y_1-y_0)$ and $P_2=(y_1+y_0+x)$. So the hypothesis of
$\mu$ is
\[
\mu_{\rm
hyp}=(y_1^2+(2y_1-1)(x+y_0)-x)(y_1+y_0+x)^{X1}(y_1^2-2y_1-y_0)^{X2}.
\]
After  substitution of this $\mu_{\rm hyp}$ into the PDE system
for the integrating factor, we obtain that $X2=-(X1+2)$ so an
integrating factor of given ODE is
\[
\mu=(y'^2+(2y'-1)(x+y)-x)(y'+y+x)^{X1}(y'^2-2y'-y)^{-(X1+2)},
\]
where $X1$ is an \emph{arbitrary} constant.
\end{example*}

Examinations of the cases when $a_k=1$ and $b_k>0$, are more
dif\/f\/icult. We are not able to give here the full analysis of
the problem which necessitates  consideration of many subcases.
\emph{For brevity}, we will only demonstrate that the method of
resultants in principle enables us to obtain all the needed
polynomials in a f\/inite number of steps. The main tool here is
usage of \emph{repeated resultants}.

For compactness, we will use the following notation for the
repeated resultants
\[R_{z_1,z_2}(f,(g,h))=R_{z_2}(R_{z_1}(f,h),R_{z_1}(g,h)).\]

For the case $a_k=1$, we obtain the following equation similar to
(\ref{C1}):
\begin{gather*}
L_4P_k =\prod_{\substack{i\\i\neq k}} \tilde{P}_i\prod_j
\tilde{Q}_j(M_1 A+M_2 B)\\
\phantom{L_4P_k =}{} + \prod_{\substack{i\\i\neq k}}
\tilde{P}_i^2\prod_j \tilde{Q}_j^2\left[\left[2\frac{\partial
A}{\partial y_{n-1}}+D(B)\right]\frac{\partial P_k}{\partial
y_{n-1}}+D(P_k)\frac{\partial B}{\partial y_{n-1}}\right], 
\end{gather*}
where $L_4$, $M_1$, $M_2$ are some polynomials, and $\tilde{P}_i$
$(i\neq k)$, $\tilde{Q}_j$ are only such polynomials from the
integrating factor structure which \emph{do not satisfy}
conditions (\ref{C12}) and (\ref{C22}). So for this case we can
obtain the following properties ($ \frac{\partial P_k}{\partial
y_{n-1}}\neq 0$):
\begin{gather}
R_{z_1,z_2}(P_k,(A,B))\mid
R_{z_1,z_2}\Bigg(\prod_{\substack{i\\i\neq k}}
\tilde{P}_i^2\prod_j \tilde{Q}_j^2\Bigg[\Bigg[2\frac{\partial
A}{\partial
y_{n-1}}+D(B)\Bigg]\frac{\partial P_k}{\partial y_{n-1}}\nonumber\\
\phantom{R_{z_1,z_2}(P_k,(A,B))\mid R_{z_1,z_2}}{}
+D(P_k)\frac{\partial B}{\partial y_{n-1}}\Bigg],(A,B)\Bigg)
\label{C4}
\end{gather}
and further
\begin{gather}
R_{z_1,z_2,z_3}\left(P_k,\left(A,B,\frac{\partial B}{\partial
y_{n-1}}\right)\right)\mid
 R_{z_1,z_2,z_3}\Bigg(\prod_{\substack{i\\i\neq k}}
\tilde{P}_i^2\prod_j \tilde{Q}_j^2\frac{\partial P_k}{\partial
y_{n-1}}\Bigg[2\frac{\partial A}{\partial
y_{n-1}}\nonumber\\
\phantom{R_{z_1,z_2,z_3}\left(P_k,\left(A,B,\frac{\partial
B}{\partial y_{n-1}}\right)\right)\mid
 R_{z_1,z_2,z_3}}{} +D(B)\Bigg],\left(A,B,\frac{\partial B}{\partial y_{n-1}}\right)\Bigg).
\label{C5}
\end{gather}

From (\ref{C4}) we can obtain the hypothesis for $P_k$ and then
with the help of (\ref{C5}) all its undetermined constants.

If there is a polynomial $P_s$ in the integrating factor structure
which satisfy (\ref{C12}), then f\/inding~$P_k$ can be
simplif\/ied. Since
\begin{gather}
R_z(P_k,P_s) \mid  R_z\Bigg(A\prod_{\substack{i\\i\neq k}}
\tilde{P}_i^2\prod_j \tilde{Q}_j^2\frac{\partial P_s}{\partial
y_{n-1}}\left[A\frac{\partial P_k}{\partial y_{n-1}}
+BD(P_k)\right],P_s\Bigg), \label{C6}
\end{gather}
then
\begin{gather}
R_{z_1,z_2}(P_k,(P_s,B))\mid
R_{z_1,z_2}\Bigg(A^2\prod_{\substack{i\\i\neq k}}
\tilde{P}_i^2\prod_j \tilde{Q}_j^2\frac{\partial P_s}{\partial
y_{n-1}}\frac{\partial P_k}{\partial y_{n-1}} ,(P_s,B)\Bigg).
\label{C7}
\end{gather}

Among other things we can observe from (\ref{C4}) and (\ref{C6})
is that it is very likely that the $P_k$'s satisfy the relation
(\ref{C12}) even if $a_k=1$. Also, there are relations for $Q_k$
with $b_k>0$, similar to (\ref{C4})--(\ref{C7}).

The main elements of the method described above were implemented
in demonstration procedures in \emph{Maple}
\cite{Kosovtsov,Kosovtsov1} for the case of rational integrating
factors.

We have tested routines in the following area: for a given ODE
order we assigned the f\/irst integral $\zeta$ of the following
type
\begin{gather*}
\zeta=\frac{\prod_i N_i^{\alpha_i}}{\prod_i M_i^{\beta_i}}
+\sum_i(\eta_i \ln \Phi_i+\theta_i \arctan \Psi_i),
\end{gather*}
where $N_i$, $M_i$, $\Phi_i$, $\Psi_i$ are some polynomials,
$\alpha_i>0$, $\beta_i>0$, $\eta_i$, $\theta_i$ are some
constants. Such algebraic f\/irst integral leads to an ODE of type
(\ref{ode}) with $f=-D(\zeta)/\frac{\partial \zeta_k}{\partial
y_{n-1}}$, which is potentially tractable by the proposed
procedure as this ODE has a rational integrating factor
$\mu=\frac{\partial \zeta_k}{\partial y_{n-1}}=\prod_i P_i^{a_i}$.

The f\/irst stage -- f\/inding candidates and hypothesis of the
structure of an integrating factor -- is very fast. The second
stage is relatively resource consuming -- f\/inding the unknown
parameters~$a_i$ of the structure of an integrating factor.

We can conclude that our prototypes are workable for rational ODEs
even with symbolic constant parameters of orders from $n=1$ to
$n=4-5$. Although in principle the method provides ability to
f\/ind an integrating factor for almost any of the above mentioned
type of ODEs, in practice the procedure heavily relies on some
basic CAS functions such as \emph{factor, simplify} and so on, and
in some cases fails. In addition, dif\/ferent outputs are observed
in dif\/ferent \emph{Maple} versions for the same ODE.

As a rule, \emph {Maple} cannot f\/ind any integrating factor for
ODEs from the considered test area while proposed procedure is
usually successful. Often our procedure produces more than one
integrating factor.  For example, the output may contain a set of
$\mu$'s or $\mu$ with arbitrary parameters, and sometimes they
lead to independent f\/irst integrals.

\section{Conclusion}

We have proposed an algebraic method for f\/inding polynomials
$P_i$ and $Q_j$ of the integrating factor form (\ref{mu1}) of
rational ODE (\ref{ode2}) for any order using resultants of the
numerator and denominator of the ODE's right-hand-side.

The main advantage of the method of resultants is ability to
obtain the polynomials $P_i$ and~$Q_j$ without an a priori
assumption of the upper bound for their respective degrees. Our
method bypasses the principal weakness of the Prelle--Singer
(Darboux) method.

Moreover, the number of undetermined variables involved in
calculations by our method is radically smaller than in the
Prelle--Singer (Darboux) method. This property allows us to obtain
polynomials $P_i$ and $Q_j$ of arbitrary degrees without any
complications whereas the Prelle--Singer (Darboux) method can not
be useful as a rule for $N>4$.

\subsection*{Acknowledgements}

I would like to thank the referees for extensive comments and
suggestions regarding of earlier versions of this paper and Reece
Heineke for a careful reading of the paper.

\LastPageEnding

\end{document}